\documentclass[journal]{IEEEtran}

\ifCLASSINFOpdf
\else
\fi

\usepackage{psfrag,epsfig}
\usepackage{amsmath,amssymb}
\usepackage{algorithm,algorithmic}
\usepackage{subfig}
\usepackage{tabularx}

\newcommand{\be}{\begin{equation}}
\newcommand{\ee}{\end{equation}}
\newcommand{\beqy}{\begin{eqnarray}}
\newcommand{\eeqy}{\end{eqnarray}}
\newcommand{\beqynn}{\begin{eqnarray*}}
\newcommand{\eeqynn}{\end{eqnarray*}}

\newcommand{\ba}{\begin{array}}
\newcommand{\ea}{\end{array}}
\newcommand{\bmx}{\begin{bmatrix}}
\newcommand{\emx}{\end{bmatrix}}
\newcommand{\bsmx}{\left[\begin{smallmatrix}}
\newcommand{\esmx}{\end{smallmatrix}\right]}
\newcommand{\bmxc}[1]{\left[\begin{array}{@{}#1@{}}}
\newcommand{\emxc}{\end{array}\right]}

\newcommand{\bt}[1]{\begin{tabular}{#1}}
\newcommand{\et}{\end{tabular}}

\newcommand{\bc}{\begin{center}}
\newcommand{\ec}{\end{center}}
\newcommand{\ben}{\begin{enumerate}}
\newcommand{\een}{\end{enumerate}}
\newcommand{\bi}{\begin{itemize}}
\newcommand{\ei}{\end{itemize}}

\newcommand{\F}{\boldsymbol{F}}
\newcommand{\G}{\boldsymbol{G}}

\newcommand{\I}{\boldsymbol{I}}

\newcommand{\K}{\boldsymbol{K}}
\renewcommand{\L}{\boldsymbol{L}}

\renewcommand{\P}{\boldsymbol{P}}
\newcommand{\Q}{\boldsymbol{Q}}
\newcommand{\R}{\boldsymbol{R}}

\newcommand{\T}{\boldsymbol{T}}
\newcommand{\U}{\boldsymbol{U}}

\newcommand{\W}{\boldsymbol{W}}

\renewcommand{\a}{\boldsymbol{a}}

\newcommand{\e}{\boldsymbol{e}}

\newcommand{\h}{\boldsymbol{h}}

\newcommand{\n}{\boldsymbol{n}}

\newcommand{\q}{\boldsymbol{q}}
\renewcommand{\r}{\boldsymbol{r}}
\newcommand{\s}{\boldsymbol{s}}

\renewcommand{\u}{\boldsymbol{u}}
\renewcommand{\v}{{\boldsymbol{v}}}
\newcommand{\w}{{\boldsymbol{w}}}
\newcommand{\x}{{\boldsymbol{x}}}
\newcommand{\y}{{\boldsymbol{y}}}
\newcommand{\z}{{\boldsymbol{z}}}

\newcommand{\0}{{\boldsymbol{0}}}

\begin{document}
%
\title{Mobile Localization in Non-Line-of-Sight Using Constrained Square-Root Unscented Kalman Filter}
%
%
%

\author{Siamak~Yousefi,~\IEEEmembership{Student Member,~IEEE,}
        Xiao-Wen~Chang,
        and~Benoit~Champagne,~\IEEEmembership{Senior Member,~IEEE}
\thanks{S. Yousefi and B. Champagne are with the Department
of Electrical and Computer Engineering, McGill University, Montreal,
QC, H3A 0E9 Canada e-mail:(siamak.yousefi@mail.mcgill.ca;benoit.champagne@mcgill.ca).}
\thanks{X.W. Chang is with the School of Computer Science, McGill University, Montreal,
QC, H3A 0E9, Canada e-mail:(chang@cs.mcgill.ca) }
}

\markboth{IEEE Transaction on Vehicular Technology (Draft)}%
{Shell \MakeLowercase{\textit{et al.}}: Bare Demo of IEEEtran.cls for Journals}
%



\maketitle

\begin{abstract}
Localization and tracking of a mobile node (MN) in non-line-of-sight (NLOS) scenarios, based on time of arrival (TOA) measurements, is considered in this work.
To this end, we develop a constrained form of square root unscented Kalman filter (SRUKF), where the sigma points of the unscented transformation are projected onto the feasible region by solving constrained optimization problems.
The feasible region is the intersection of several discs formed by the NLOS measurements.
We show how we can reduce the size of the optimization problem and formulate it as a convex quadratically constrained quadratic program (QCQP), which depends on the Cholesky factor of the \textit{a posteriori} error covariance matrix of SRUKF.
As a result of these modifications, the proposed constrained SRUKF (CSRUKF) is more efficient and has better numerical stability compared to the constrained UKF.
Through simulations, we also show that the CSRUKF achieves a smaller localization error compared to other techniques and that its performance is robust under different NLOS conditions.
\end{abstract}


%
\IEEEpeerreviewmaketitle

\section{Introduction}
\IEEEPARstart{N}{etwork}-based radio localization has received great attention in recent years due to limitations of the global positioning system (GPS) in indoor places and dense urban areas, and finds numerous applications in surveillance, security, etc. \cite{Sayed}.
In this technology, radio signals exchanged between a mobile node (MN) and fixed reference nodes (RN) with known positions
\footnote{In wireless cellular networks, the RN are identified with the base stations, while in wireless sensor networks, they are called anchors.}, are exploited to determine the unknown location of the MN.
There are several different measurements that can be used for localization, e.g., time of arrival (TOA), time difference of arrival (TDOA), received signal strength (RSS), angle of arrival (AOA), and a hybrid of these data.
Amongst different localization techniques, TOA-based methods in which the MN and RNs are synchronized are usually preferred, especially in the context of IEEE 802.15.4a which exploits ultra band-with (UWB) technology \cite{IEEE_802.15.4.a}.
Indeed, measurement of TOA can be done accurately with UWB signaling due to its fine timing resolution and robustness against multipath and fading.


One of the main challenges in radio localization is the non-line of sight (NLOS) problem, which occurs due to the blockage of the direct sight between the MN and RNs.
In a NLOS situation, either due to reflection of the radio waves through scatterers or penetration through blocking objects, the travel time of the received signals increases  \cite{Alsindi-Pahlavan}, \cite{Identification_Buehrer}, \cite{Identify-UWB-Marano}.
Consequently, the NLOS error of each measured TOA needs to be modelled as a random variable with a positive bias, which can be quite large \cite{Guvenc}.
The first step in dealing with the NLOS problem is to detect the NLOS measurements and, if necessary, to discard them.
To this end, some techniques estimate the variance of each measurement, and if it is above a given threshold, the corresponding link is identified as NLOS \cite{Thomas_TDOA_AOA}, \cite{EKF_TDOA_AOA}, \cite{Yu_NLOS_Identification_Mitigation}.
NLOS identification techniques using signal features have also been proposed for UWB applications \cite{Identification_Buehrer,Identify-UWB-Guvenc,Identify-UWB-Marano}.
If, however, the NLOS measurements cannot be discarded due to an insufficient number of LOS measurements for unambiguous localization, the next step is to mitigate their effect through further processing.

There are numerous works focusing on NLOS mitigation for the localization of stationary nodes, which are mostly based on (memoryless) constrained optimization techniques.
In these approaches, the position of the MN is constrained to be within the convex hull formed by the intersection of multiple discs, each disc being centered at one of the NLOS RNs and with a radius equal to the corresponding measured range.
By restricting the MN position in this way and by employing the LOS measurements in the cost function, the unknown location can be found through solving a constrained optimization problem.
For a survey on TOA-based memoryless localization in NLOS scenarios, see \cite{Guvenc} and the references therein.

For an MN with available dynamic model, filtering techniques are preferred compared to memoryless methods.
This is especially the case when data from inertial measurements units (IMU) are used in parallel with range information for tracking purposes \cite{Tightly_Coupled_UWB}, \cite{Loosely_Coupled_Youssef}.
Some methods apply Kalman filter preprocessing on measured TOAs to smooth out the effect of the variances of the NLOS biases, while scaling the covariance matrix in an extended Kalman filter (EKF) to further mitigate their means
\cite{Thomas_TDOA_AOA}, \cite{EKF_TDOA_AOA}, \cite{Yu_NLOS_Identification_Mitigation}.
However, these approaches can only achieve a moderate performance for large NLOS biases.
In \cite{IMM_Chen,IMM_Fritsche}, it is assumed that the mean and variance of the NLOS biases are known; in practive, however, this information is not available accurately beforehand unless prior field measurements are obtained.

Some other approaches regard the NLOS bias as a nuisance parameter and try to estimate its distribution using Kernel density estimation (KDE).
In \cite{Hammes1}, a robust semi-parametric EKF is proposed for NLOS mitigation of an MN.
The performance of this technique is improved by the interacting multiple model (IMM) algorithm in \cite{IMM_Hammes}.
Although considered for TOA measurements, these techniques are also suitable when AOA, RSS or a hybrid of these are employed.
However, in addition to high computational cost, the performance of KDE still depends on how well it can model the distribution of the NLOS biases.
It is claimed that for cellular applications, the performance is only satisfactory when the ratio of NLOS to LOS measurements is less than a half and a higher ratio might result in divergence of KDE algorithms \cite{IMM_Hammes}.

In some other techniques, the random NLOS biases are considered as parameters in the state vector, to be jointly estimated with other state parameters \cite{EKF_Najar,Jourdan,Gonzalez,Siamak1}, while the NLOS bias variation over time is modelled as a random walk.
The technique in \cite{EKF_Najar} uses EKF, while \cite{Jourdan} and \cite{Gonzalez} use particle filters (PFs) that generally have a high computational cost.
In \cite{Siamak1}, an improved EKF is used where bound constraints on the NLOS biases are
enforced for improving the localization accuracy.
Although the above techniques can mitigate the effect of NLOS biases to some extent,
their performances might not be good due to the mismatch between the random walk model and the physical reality, which is unavoidable considering the unpredictable nature of the biases.

In this work, we propose an efficient square root unscented Kalman filter (SRUKF) with convex inequality constraints for localization of an MN in NLOS situation.
The proposed constrained SRUKF (CSRUKF) is mainly based on a combination of the SRUKF in \cite{SRUKF} for unconstrained problems and the constrained UKF in \cite{UT_Projection_Kandepu}.
In our proposed algorithm, similar to some memoryless approaches, the NLOS measurements are removed from the observation vector and are employed instead to form a closed convex constraint region \cite{Guvenc}.
At each time step, we use a SRUKF to estimate the state vector and compute the Cholesky factor of the error covariance matrix.
To impose the constraints onto the estimated quantities, as proposed in \cite{UT_Projection_Kandepu}, the sigma points of the unscented transformation may need to be projected onto the feasible region by solving a convex quadratically constrained quadratic program (QCQP).
However, we show that the projection can be done in a more efficient and numerically stable way by solving a QCQP with reduced size, in which the cost function depends on the Cholesky factor of the \textit{a posteriori} error covariance matrix, readily obtained from the SRUKF.
Through simulations, our proposed algorithm is shown to achieve a good localization performance under different NLOS scenarios.
In particular, in severe NLOS conditions, i.e., when there is only one or no LOS measurement, our method achieves a superior performance compared to other benchmark approaches.
Another positive aspect is its robustness to false alarm (FA) errors in NLOS identification, which makes it suitable for practical applications where such errors may be inevitable.


%

The organization of the paper is as follows:
In Section \ref{Sec:System}, the system model is described and the problem formulation is presented.
The proposed constrained SRUKF algorithm is developed in Section \ref{Sec:UT}, along with a discussion of computational complexity.
The simulation results and comparisons with different algorithms are given in Section \ref{Sec:Simulation}.
Finally, Section \ref{Sec:Conclusion} concludes the paper.

\textit{Notation:}
Small and capital bold letters represent vectors and matrices respectively.
The vector 2-norm operation is denoted by $\|\cdot\|$, and $(\cdot)^T$ and $(\cdot)^{-1}$ stand for matrix transpose and inverse operations, respectively.
A diagonal matrix with entries $x_1,\ldots,x_M$ on the main diagonal is denoted by $\textrm{diag}(x_1,\ldots,x_M)$.
For $i \leq j$, $\q(i\!:\!j)$ denotes a vector of size $j-i+1$ obtained by extracting the $i$-th to $j$-th entries of vector $\q$, inclusively.
The symbol $\I$ denotes an identity matrix of appropriate dimension.
For a positive semi-definite Hermitian matrix $\R$, $\R^{1/2}$ denotes its unique positive semi-definite definite square root, i.e. such that $\R^{1/2}\R^{1/2}=\R$ \cite{Horn}.

%
%
%
%
%
%

\section{System Description and Problem Statement} \label{Sec:System}

\subsection{System Model}
Consider a network of $M$ fixed RNs and one MN, distributed on a 2-dimensional (2D) plane and exchanging timing signals via wireless links.
With reference to a Cartesian coordinate system in this plane,
let $\a^{i} \in \mathbb{R}^2$ denote the known position vector of the $i$-th RN, where $i \in \{1,\ldots,M\}$,
while $\x_k \in \mathbb{R}^2$ and $\v_k \in \mathbb{R}^2$ denote the unknown position and velocity vectors of the MN at discrete time instant $k$, respectively.
Let the state vector be $\s_k =[\x_k^T,\v_k^T ]^T \in \mathbb{R}^4$, which includes the position and velocity components of the MN.
The motion model is assumed to be
\be \label{State}
\s_{k} = \F \s_{k-1} + \G \w_{k-1},
\ee
where the matrices $\F$ and $\G$ are
\be
\F =
\bmx
1 & 0 & \delta t & 0 \\
0 & 1 & 0 & \delta t \\
0 & 0 & 1 & 0\\
0 & 0 & 0 & 1 \\
\emx ,
~ \quad
\G =
\bmx
\frac{\delta t^2}{2}  & 0 \\
0 & \frac{\delta t^2}{2} \\
\delta t & 0 \\
 0 & \delta t\\
\emx ,
\ee
and $\delta t$ is the time step duration.
The vector $\w_{k-1} \in \mathbb{R}^2$ in \eqref{State} is a zero-mean white Gaussian noise process
with diagonal covariance matrix $\Q= \sigma_w^2 \I$.

In this work, we consider TOA-bsed localization, in which the range between the MN and each RN is obtained by multiplying the time of flight of the radio wave by the speed of light.
If the MN and RNs are accurately synchronized, then a one-way ranging scheme can be used;
otherwise, a two-way ranging protocol may be employed
where the relative clock offsets are removed from the TOA measurements \cite{Dardari}.
Let ${\cal L}_k $ and ${\cal N}_k$ denote the index sets of the RNs that are identified as LOS and NLOS nodes at time instant $k$, respectively.
The range measurements can thus be represented as
\be \label{Range_Measurement}
r_k^{i} =
\begin{cases}
                h^i(\s_k) + n_{k}^{i},     & i \in {\cal L}_k, \\
                h^i(\s_k) + b_{k}^{i} + n_{k}^{i},   & i \in {\cal N}_k, \\
\end{cases}
\ee
where $h^i(\s_k)=\| \x_k - \a^{i} \|$, $n_k^{i}$ is the measurement noise and $b_k^i$ is a positive random NLOS bias, which is usually considered independent from $n_k^i$.
The noise terms $n_k^i$, for $i=1,\ldots,M$, are modelled as independent white Gaussian processes, with zero-mean and known variance $\sigma_n^2$.
The distribution of each NLOS biases $b_k^i$ is time-varying due to the movement of the MN and other objects in the area.
In the literature, different distributions have been considered for these biases, for instance:
exponential \cite{Molisch}, \cite{Yu_SQP},
shifted Gaussian \cite{IMM_Fritsche},
and uniform \cite{Identification_Buehrer} are widely employed.
However, having \emph{a priori} knowledge about the distribution of the NLOS biases requires preliminary field measurements, which may not be possible in practical applications.
Therefore, in this work, we do not make any specific assumption about the distribution of the NLOS biases, although we suppose that the NLOS links are identified accurately at every time instant.

\subsection{Problem Formulation}
The state vector $\s_k$ and the NLOS biases $b_k^{i}$ for $i \in {\cal N}_k$ are the unknown parameters in the above model.
Representing the NLOS biases by a simple dynamic model such as a random walk, as considered in \cite{Jourdan}, \cite{Gonzalez}, may not be an accurate approximation, and thus this approach is not used in this work.
To simplify the problem and reduce the number of unknowns, we eliminate the NLOS measurements from the observation vector, and instead use the information carried out by the biases to restrict the position of the MN within a certain range.
For instance, in many applications, it can be assumed that the TOA measurement noise $n_{k}^{i}$ is small compared to $b_k^{i}$ (especially in UWB ranging), which implies that $b_k^i + n_k^i \geq 0 $  \cite{Guvenc}.
In light of (3), this assumption is equivalent to
\be \label{Feasible_Region_NLOS}
\|\x_k-\a^i \| \leq r_k^i, \quad i \in {\cal N}_k,
\ee
which is obviously a convex constraint as in \cite{Doherty_SDP}.
If the small noise assumption cannot be made, e.g., in narrowband systems where TOA-based ranging measurement errors are relatively large, the constraints in \eqref{Feasible_Region_NLOS} may not be satisfied.
To avoid this limitation, we can generalize the latter as
\be
\label{Feasible_Region_NLOS_2}
\|\x_k-\a^i \| \leq r_k^i + \epsilon \sigma_n, \quad i \in {\cal N}_k,
\ee
where $\epsilon \geq 0$ is a small number to ensure that the MN is located inside a disc with radius $r_k^i+ \epsilon \sigma_n$.
Note that even if the bias is zero, i.e., in LOS situation, it is more likely that the MN satisfies the constraint in \eqref{Feasible_Region_NLOS_2} as compared to \eqref{Feasible_Region_NLOS}.
Therefore, we propose to use the constraint in \eqref{Feasible_Region_NLOS_2} throughout this work due to its robustness against measurement noise and false alarm (FA) error in NLOS identification.
In the sequel, the \textit{feasible region} refers to the convex feasible set formed by the intersection of the discs in \eqref{Feasible_Region_NLOS_2}, denoted by ${\cal D}_k$, hence
\be
{\cal D}_k = \Big \lbrace \x: \| \x - \a^i \| \leq r_k^i + \epsilon \sigma_n, \forall i \in {\cal N}_k \Big \rbrace \nonumber.
\ee

At every time instant $k$, let us remove the NLOS measurements from the observations in \eqref{Range_Measurement} and only keep the LOS measurements, i.e., $ r_k^i$ for all $i \in {\cal L}_k $.
The remaining LOS range measurements can be represented by the vector  $\z_k \in \mathbb{R}^{|{\cal L}_k |}$.
Note that in the worst case, where all the measurements are identified as NLOS, the vector $\z_k$ is empty.
The state space model and constraints can thus be expressed as
\begin{subequations}
\begin{align} \label{Measurement_Process2_a}
&\z_k = \h(\s_k) + \n_k, \\
\label{Measurement_Process2_b}
&\s_{k} = \F \s_{k-1} + \G \w_{k-1}, \\
\label{Measurement_Process2_c}
& \|\x_k-\a^i \| \leq r_k^i + \epsilon \sigma_n, \quad i \in {\cal N}_k,
\end{align}
\end{subequations}
where $\h(\s_k)$ and $\n_k$ are vectors whose entries are $h^i(\s_k)$ and $n_k^i$ for every $i \in {\cal L}_k$, respectively.
Under our previous assumptions on the measurement noise $n_k^i$ in (3), the covariance matrix of
$\n_k$ is positive-definite diagonal, i.e. $\R=\mathbb{E}[ \n_k \n_k^T ]=\sigma_n^2 \I \in \mathbb{R}^{ |{\cal L}_k| \times  |{\cal L}_k| }$.
The constraints in \eqref{Measurement_Process2_c} are only on the first two elements of the state vector, i.e., $\x_k$, as we have a 2D positioning scenario herein.
Note that if the constraints in \eqref{Measurement_Process2_c} are removed from the state model, then an ordinary nonlinear filtering technique such as EKF can be used.
This approach is also known as EKF with outlier rejection since the NLOS measurements are regarded as outliers and therefore discarded.

In minimum mean square error (MMSE) estimation, e.g., Kalman-type filters, one tries to find the conditional mean and covariance matrix of the state vector $\s_k$ given the measurements up to current time instant $k$, with the conditional probability density function (PDF) $f( \s_k | \z_1, \ldots , \z_k )$.
However, when extra information about the state vector is available in the form of inequality constraints, the probability that the MN is outside the feasible region should be zero. 
Hence a truncated or \emph{constrained} conditional PDF, $f_c(.|.)$, can be defined as
\be
f_c( \s_k | \z_1, \ldots , \z_k ) =
\begin{cases}
\frac{1}{\beta} f( \s_k | \z_1, \ldots , \z_k ), \quad & \textrm{if} ~\x_k \in {\cal D}_k, \\
0          ,                                      \quad & \textrm{otherwise},
\end{cases}
\ee
where
$\beta \triangleq \int_{ \x_k \in {\cal D}_k} f(\s_k | \z_1, \ldots , \z_k ) d\s_k$
is a normalization constant.
Therefore, one can estimate the state vector by finding the conditional mean of $\s_k$ with truncated PDF as
\be \label{Constrained_ML}
\hat{\s}_k =  \int_{\x_k \in {\cal D}_k} \s_k f_c( \s_k | \z_1, \ldots , \z_k ) d\s_k,
\ee
and the covariance matrix of the constrained state estimate can be found through
\be \label{Constrained_Covariance}
\hat{\boldsymbol \Sigma}_{k} = \int_{\x_k \in {\cal D}_k} (\s_k- \hat{\s}_k) (\s_k- \hat{\s}_k )^T f_c( \s_k | \z_1, \ldots , \z_k ) d \s_k.
\ee
%
This idea is known as PDF truncation, where the distribution of the state vector given the measurements
is forced to be zero outside the feasible region \cite{Simon_Survey}.
For a linear dynamic model with zero-mean Gaussian measurement and process noises, where the state vector is subject to linear inequality constraints, closed form expressions for 
$\hat{\s}_k$ and $\hat{\boldsymbol \Sigma}_{k}$ in \eqref{Constrained_ML}-\eqref{Constrained_Covariance} have been obtained using PDF truncation along with the Gaussian assumption \cite{Simon_Constrained_KF_Truncation}.
For nonlinear inequality constraints, it is proposed to do a Taylor series linearization of the constraints around the current state estimate and then apply the aforementioned method; however, this approach may not be accurate \cite{UT_Projection_Lan}.
In general cases with nonlinear inequality constraints, PDF truncation requires multidimensional Monte Carlo integration which becomes computationally expensive as the size of the state vector grows.
Therefore, PDF truncation may not be a good approach to solve our problem.
In the following section, we show how we can efficiently approximate $\hat{\s}_k$ and $\hat{\boldsymbol \Sigma}_{k}$ using an alternative approach that combines the SKURF [21] for unconstrained problems with the projection-based constrained UKF in [22].

\section{Constrained Nonlinear Filtering with Sigma Point Projection}\label{Sec:UT}
Another family of methods for imposing inequality constraints on the state vector are the projection-based techniques, in which the unconstrained state estimate, obtained through a Kalman-type filter, is projected onto the feasible region by solving an optimization problem \cite{Simon_Survey}.
However, by this approach, one cannot estimate the constrained error covariance matrix of the state, i.e., $\hat{\boldsymbol \Sigma}_{k}$, accurately.
Therefore, in addition to the unconstrained state estimate, some representative sample points of the conditional PDF $f(\s_k|\z_1,\ldots,\z_k)$  need to be projected onto the feasible region.
For instance, the sigma points of the unscented transformation (UT) can give good statistical information about the mean and the error covariance matrix of the state estimate \cite{Julier_UT}.
Based on this idea, in \cite{UT_Projection_Kandepu}, a constrained UKF technique has been proposed in which the sigma points of the UKF violating the constraints are projected onto the feasible region.
However, due to the dependence of the projection function on the inverse of the \textit{a posteriori} error covariance matrix, the method in \cite{UT_Projection_Kandepu} may become numerically unstable \cite{UT_Projection_Lan}.
In the following subsections, we first describe a variation of the SRUKF that is better suited to our specific problem; then, to overcome the above mentioned numerical issue, we design a more efficient and numerically reliable method for projecting the sigma points generated from the \emph{a posteriori} estimates, onto the feasible region; finally, we summarize our algorithm and comment on its numerical complexity.

\subsection{Unconstrained SRUKF Algorithm}
The proposed algorithm in this part is based on the SRUKF presented in \cite{SRUKF} with slight modification such that the algorithm is more efficient and numerically reliable.
Let $\s_{k-1|k-1}$ be the estimated state and $ \boldsymbol \Sigma_{k-1|k-1}$ be the estimated error covariance matrix of the state, based on the available measurements up to current time instant $k-1$.
Let $\U_{k-1|k-1}$ be the upper triangular Cholesky factor of $ \boldsymbol \Sigma_{k-1|k-1}$, i.e., $ \boldsymbol \Sigma_{k-1|k-1} = \U_{k-1|k-1}^T \U_{k-1|k-1}$.
Then, for the next time instant, the \textit{a priori} estimate of the state vector and the corresponding error covariance matrix, denoted as $\s_{k|k-1}$ and $\boldsymbol \Sigma_{k|k-1}$, respectively, can be obtained through prediction as
\begin{align}
\label{Prediction_Mean}
\s_{k|k-1} &= \F \s_{k-1|k-1} ,\\
\label{Prediction_Covariance}
\boldsymbol \Sigma_{k|k-1} &= \F \boldsymbol \Sigma_{k-1|k-1} \F^T+ \G \Q \G^T .
\end{align}
Alternatively, the computation of \eqref{Prediction_Covariance} can be avoided as only the Cholesky factor of the \textit{a priori} covariance matrix, denoted by $\U_{k|k-1}$ is required \cite{SRUKF}.
To this aim, let us rewrite \eqref{Prediction_Covariance} as
\be \label{Covariance_Factorization}
\boldsymbol \Sigma_{k|k-1} =
\begin{bmatrix}
\F \U_{k-1|k-1}^T & \G \Q^{\frac{1}{2}}
\end{bmatrix}
\begin{bmatrix}
\U_{k-1|k-1} \F^T \\
\Q^{\frac{1}{2}} \G^T
\end{bmatrix} ,
\ee

If we compute the QR factorization of the second matrix on the right hand side of  \eqref{Covariance_Factorization},
we obtain $\U_{k|k-1}$:
\be \label{Prediction_Covariance_Cholesky}
\U_{k|k-1} = \textrm{qr} \left\lbrace
\begin{bmatrix}
\U_{k-1|k-1} \F^T \\
\Q^{\frac{1}{2}} \G^T
\end{bmatrix} \right\rbrace ,
\ee
where by definition, the function \textrm{qr}\{.\} returns the upper triangular factor of the QR factorization of its matrix argument.

With the help of $U_{k|k-1}$,
the sigma points of the SRUKF are generated as proposed in \cite{SRUKF}, i.e.:
\be \label{Sigma_points_Unconstrained}
\s_{k|k-1}^{(j)} = \begin{cases}
              \s_{k|k-1}  ,             & j = 0,  \\
              \s_{k|k-1}+\sqrt{\eta_{\alpha}} (\U_{k|k-1}^T)_j  , & j = 1, \ldots, N, \\
              \s_{k|k-1}-\sqrt{\eta_{\alpha} }  (\U_{k|k-1}^T)_{j-N} ,  & j = N+1, \ldots, 2N , \\
           \end{cases}
\ee
where $N$ is the dimension of the state vector (in this work, $N=4$),
$(\U_{k|k-1}^T )_j$ denotes the $j$-th column of matrix $\U_{k|k-1}^T $, and 
$\eta_{\alpha}$ is a tuning parameter which controls the spread of the sigma points.
To better understand the geometric meaning of parameter $\eta_{\alpha}$, we can assume that $\s_{k|k-1}$ and
$\boldsymbol \Sigma_{k|k-1}$ obtained through the proposed filter are approximately equal to the mean and covariance matrix of the conditional PDF $f(\s_k | \z_1,\ldots, \z_{k-1})$.
Define random variable $\eta_k = (\s_k - \s_{k|k-1})^T \boldsymbol \Sigma_{k|k-1}^{-1} (\s_k - \s_{k|k-1})$, which is the weighted squared distance between $\s_k$ and $\s_{k|k-1}$.
Suppose that the parameter $\eta_\alpha$ in \eqref{Sigma_points_Unconstrained} is chosen such that
$\textrm{Pr}(\eta_k \leq \eta_\alpha) = \alpha$,
where $0 < \alpha \le 1$ represents a desired confidence level.
Then, the region of $\mathbb{R}^N$ defined by $\eta_k \le \eta_\alpha$ represents a confidence ellipsoid, on the boundary of which the sigma points in \eqref{Sigma_points_Unconstrained} fall.
For example, if $\alpha =0.9$, the probability for $\s_k$ to lie inside the ellipsoid delimited by the sigma points with the corresponding $\eta_{\alpha}$ is $90\%$.
If we assume that $f(\s_k | \z_1,\ldots, \z_{k-1})$ is approximately Gaussian, then the random variable $\eta$ has a Chi-square distribution with $N$ degrees of freedom and it becomes easy to find a value for $\eta_{\alpha}$ corresponding to a certain ellipsoid with confidence level $\alpha$
\footnote{The Matlab function $\texttt{chi2inv}(\alpha,N)$ can be used for this purpose.}.

The generated sigma points are transformed through the nonlinear measurement function as
\be \label{Measurement_Prediction}
\z_{k|k-1}^{(j)} = \h(\s_{k|k-1}^{(j)} ), \quad j=0,\ldots, 2N.
\ee
Then, the mean, cross-covariance matrix, and error covariance matrix of the transformed sigma points can be estimated by means of weighted sums as in \cite{Book_Simon}:
\begin{align} \label{Measurement_Mean}
\hat{\z}_{k|k-1}&=\sum_{j=0}^{2N}w^{(j)}\z_{k|k-1}^{(j)}, \\
\label{Cross_Covariance}
\boldsymbol \Sigma_{k|k-1}^{\s,\z}&=\sum_{j=0}^{2N}w^{(j)}(\s_{k|k-1}^{(j)}-\s_{k|k-1})(\z_{k|k-1}^{(j)}-\hat{\z}_{k|k-1})^T, \\
\label{Measurement_Cov}
\P_{k|k-1}^{z}&=\sum_{j=0}^{2N}w^{(j)}(\z_{k|k-1}^{(j)}-\hat{\z}_{k|k-1})(\cdot)^T + \R ,
\end{align}
where $(.)$ means that the same argument contained in the previous parenthesis is used, $\R$ is the covariance matrix of the measurement noise $\n_k$ in (6a) and the weights $w^{(j)}$ appearing in these expressions are defined in a similar way 
as in \cite{Zachariah}:
\begin{equation}
w^{(j)} =
\begin{cases}
 1 - \frac{N} {\eta_{\alpha}}, \quad &j=0, \\
 \frac{1} {2\eta_{\alpha} }, \quad  &j = 1, \ldots, 2N,
\end{cases}
\end{equation}
and therefore satisfy
$\sum_{j=0}^{2N} w^{(j)} = 1$.

If the weight $w^{(0)}$ in (19) is negative, it is possible that the covariance matrix obtained through \eqref{Measurement_Cov} becomes indefinite (i.e., with negative eigenvalues).
However, by choosing a sufficiently large value of $\alpha$, we can guarantee that $\eta_{\alpha} \geq N$; in turn, this implies that $w^{(0)}\geq 0$ and the covariance matrix  (17) then becomes positive definite.
In this work, we are interested in projecting the sigma points that are far away from the mean to cover a large confidence ellipsoid, and it is therefore legitimate to consider ellipsoids with larger confidence levels, so that the above issue can 
be naturally avoided \footnote{In \cite{Julier_UT}, a scaled version of unscented transformation has been proposed to capture higher moments of the nonlinear function where the generated sigma points are located in the vicinity of each other. This method also guarantees positive definiteness of the covariance matrix.
However, our problem is not highly nonlinear and we are interested to generate sigma points that might be far away from one another, therefore, our parameter selection is different from \cite{Julier_UT} and \cite{SRUKF}.}.
In our dynamic model, with $N=4$ and based on the Chi-square assumption for $\eta_k$, it follows that if $\alpha > 0.6$, then $\eta_{\alpha} > N$ and the positive definiteness of (17) is guaranteed.

For numerical stability, instead of forming $\P_{k|k-1}^z$ explicitly, its Cholesky factor is calculated \cite{SRUKF}.
Specifically, if we let
\be
\mathbf{e}_z^{(j)} = \sqrt{w^{(j)}} (\z_{k|k-1}^{(j)} -\hat{\z}_{k|k-1} ), \quad  j=0,\ldots, 2N,
\ee
then the upper triangular Cholesky factor of $\P_{k|k-1}^z$, denoted by $\U_{\z_k}$ is obtained through
\be \label{Cholesky_Measurement_QR}
\U_{\z_k} = \textrm{qr} \left\lbrace \big[ \mathbf{e}_z^{(0)},   \mathbf{e}_z^{(1)}, \ldots, \mathbf{e}_z^{(2N)}, \R^{\frac{1}{2}} \big]^T \right\rbrace .
\ee

It is proposed in \cite{SRUKF} to first compute the Kalman gain
\be \label{Kalman_Gain0}
\K_k = \boldsymbol \Sigma_{k|k-1}^{\s,\z } (\P_{k|k-1}^{z})^{-1} = \boldsymbol \Sigma_{k|k-1}^{\s,\z } \U_{z_k}^{-1} \U_{z_k}^{-T},
\ee
and then, the $\textit{a posteriori}$ state estimate and the Cholesky factor of the error covariance matrix can be updated through
\begin{align} \label{UKF_Aposterior0}
\s_{k|k} &= \s_{k|k-1} + \K_k(\z_k - \hat{\z}_{k|k-1} ),  \\
\label{UKF_Aposterior_Covariance0}
\U_{k|k} &= \textrm{cholupdate} \{ \U_{k|k-1},  \K_k  \U_{z_k}^T, -1 \} ,
\end{align}
where $\textrm{cholupdate}\{ \U_{k|k-1}, \K_k  \U_{z_k}^T, -1 \}$ is the consecutive downdates of the Cholesky factor of $\U_{k|k-1}^T\U_{k|k-1}$ using the columns of $\K_k  \U_{z_k}^T$ \cite{SRUKF}\footnote{In Matlab, the built-in function \texttt{Cholupdate} can be employed to do rank-1 Cholesky update or downdate, indicated by the third argument of the function.}, where
\be
\boldsymbol \Sigma_{k|k} \triangleq \U_{k|k}^T \U_{k|k} = \U_{k|k-1}^T \U_{k|k-1} - \K_k \U_{z_k}^T \U_{z_k} \K_k^T. \nonumber
\ee

Herein, however, we propose a more efficient and numerically reliable way to compute  $\s_{k|k}$ and $\U_{k|k}$.
Instead of the Kalman gain $\K_k$, we compute
\be \label{Kalman_Gain}
\T_k = \boldsymbol \Sigma_{k|k-1}^{\s,\z } \U_{z_k}^{-1} ,
\ee
which can be obtained by solving multiple triangular linear systems  $\T_k \U_{z,k}=\boldsymbol \Sigma_{k|k-1}^{\s,\z }$.
Then, it follows from \eqref{Kalman_Gain0} that $\K_k=\T_k  \U_{z_k}^{-T}$.
Substituting this expression into \eqref{UKF_Aposterior0} we obtain
\be \label{UKF_Aposterior}
\s_{k|k} = \s_{k|k-1} + \T_k  \U_{z_k}^{-T}(\z_k - \hat{\z}_{k|k-1} ),
\ee
where the vector $\y_k \triangleq \U_{z_k}^{-T}(\z_k - \hat{\z}_{k|k-1} )$ can be obtained by solving
the triangular linear system
$$
\U_{z_k}^T \y_k = \z_k - \hat{\z}_{k|k-1} .
$$
From \eqref{UKF_Aposterior_Covariance0} and \eqref{Kalman_Gain}, it follows that the covariance matrix can be updated as
\be
\label{UKF_Aposterior_Covariance}
\boldsymbol \Sigma_{k|k}  =  \U_{k|k-1}^T \U_{k|k-1} - \T_k    \T_k^T ,
\ee
hence the Cholesky factor of  $\boldsymbol \Sigma_{k|k}$ can be computed as
\be \label{Cholesky_Aposterior}
\U_{k|k} = \textrm{cholupdate} \{ \U_{k|k-1},  \T_k, -1 \} ,
\ee
Compared to the algorithm in \cite{SRUKF}, this modified algorithm for the estimation of $\s_{k|k}$ and $\U_{k|k}$ saves about $2N|{\cal N}_k|^2$ flops at each time step $k$.
It is also more numerically reliable as it avoids solving some linear systems, which could be ill-conditioned, and computing some matrix-matrix multiplications.

Note that if all the measurements at time instant $k$ are in NLOS, then the measurement vector $\z_k$ is empty, hence we will use the predicted state in \eqref{Prediction_Mean} and the Cholesky factor of the predicted covariance matrix in \eqref{Prediction_Covariance_Cholesky} to replace the \textit{a posteriori} state vector in \eqref{UKF_Aposterior} and Cholesky factor of the error covariance matrix in \eqref{Cholesky_Aposterior}, respectively.

\begin{figure*}[htbp]
\centering
\subfloat[] {
\includegraphics[width=55mm,height=65mm]{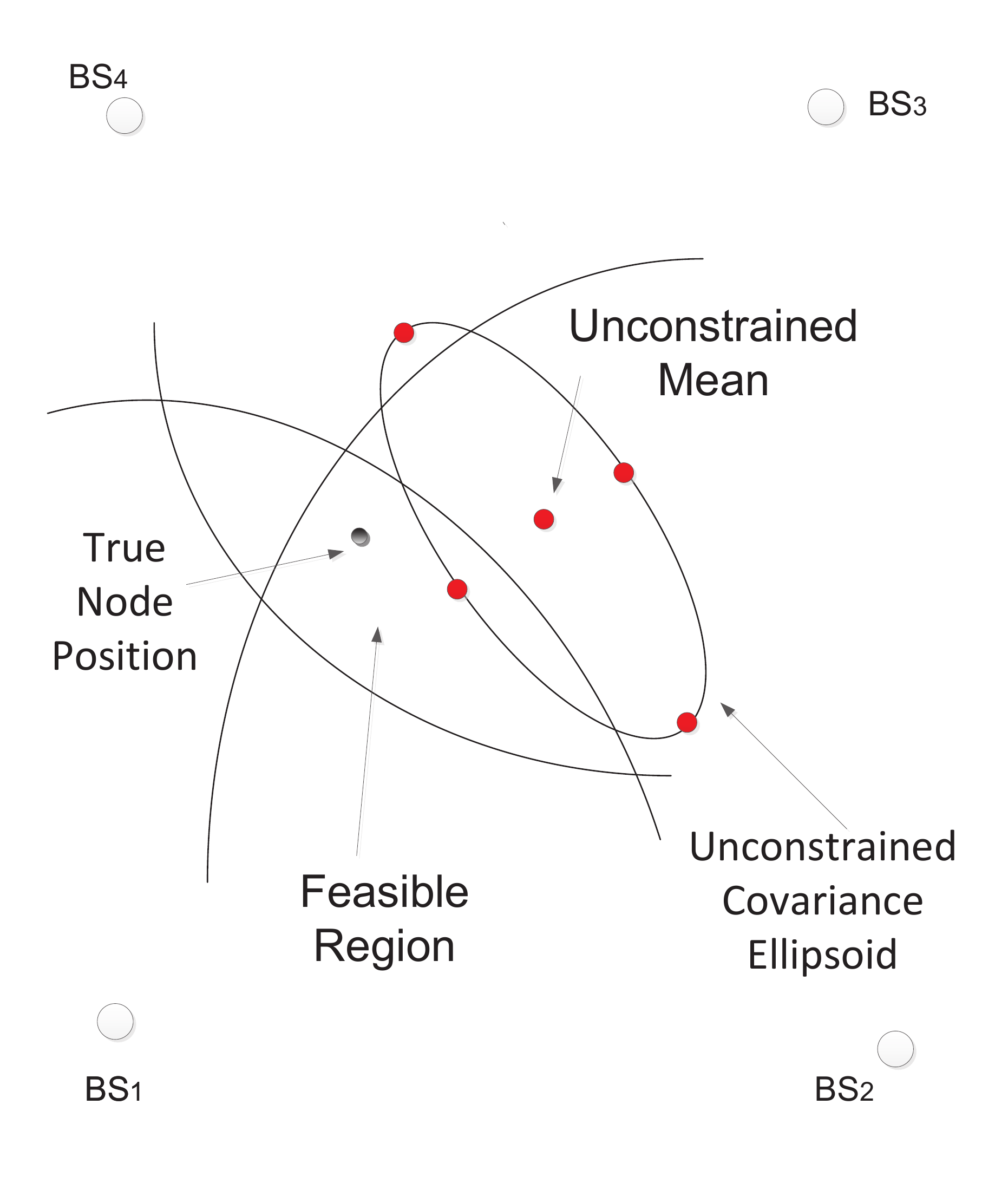}
}\quad \quad \quad
\subfloat[] {
\includegraphics[width=55mm,height=65mm]{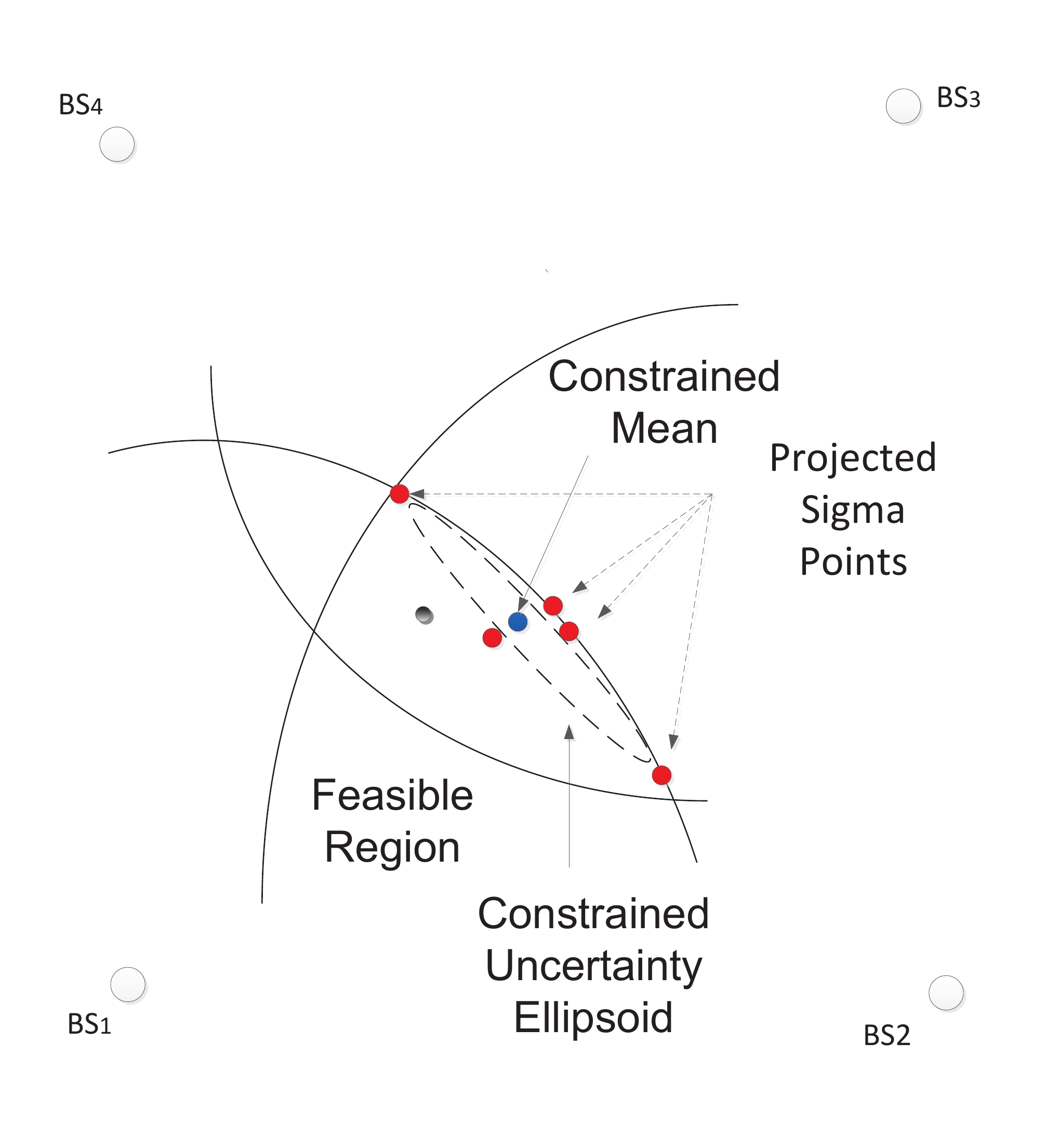}
}
\caption{Illustration of the proposed projection technique: (a) Unconstrained state estimate and the uncertainty ellipsoid of sigma points, amongst which some are outside the feasible region. (b) The projected sigma points fall inside the feasible region and the uncertainty ellipsoid is shrunk.}
\label{fig:Fig1}
\end{figure*}

\subsection{Imposing the Constraints on the Estimates} \label{Subsec:Convex}

Up to this point, the \textit{a posteriori} state estimate and the Cholesky factor of the \textit{a posteriori} error covariance matrix have been obtained using a SRUKF without taking the constraints \eqref{Measurement_Process2_c} into account.
To impose the constraints on the estimated state and error covariance matrix, similar to \cite{UT_Projection_Kandepu}, a new set of sigma points are generated according to
\begin{equation} \label{Sigma_points_Unconstrained_2}
\hspace{-.5mm}  \s_{k|k}^{(j)} = \begin{cases}
              \s_{k|k}      ,         & j = 0,  \\
              \s_{k|k}+ \sqrt{\eta_{\alpha} } (\U_{k|k}^T)_j,   & j = 1, \ldots, N, \\
              \s_{k|k}- \sqrt{\eta_{\alpha} } (\U_{k|k}^T)_{j-N} ,   & j = N+1, \ldots, 2N. \\
           \end{cases} \hspace{-1mm}
\end{equation}
The generated sigma points form an uncertainty ellipsoid with $\s_{k|k}$ at its centre as illustrated in Fig. \ref{fig:Fig1} for $N=2$.
%
After the generation of sigma points $\s_{k|k}^{(j)}$ with desired confidence ellipsoid, those which violate the constraints
are projected onto the convex feasible region through
\be \label{Projection_Sigma_Points}
\begin{split}
{\cal P}(\s_{k|k}^{(j)}) &= \mathrm{arg}\min_{\q} ~ \Big\lbrace (\q- \s_{k|k}^{(j)})^{T} \W_k (\q- \s_{k|k}^{(j)})  \Big\rbrace , \\
 \mathrm{s.t.} & \quad \big \| \q(1\!:\!2) - \a^i \big \| \leq r_k^i+ \epsilon \sigma_n, \quad i \in {\cal N}_k,
\end{split}
\ee
where $\W_k$ is a symmetric positive definite (SPD) weighting matrix \cite{UT_Projection_Lan}, \cite{Book_Simon}.
One reasonable choice is $\W_k=\boldsymbol \Sigma_{k|k}^{-1}$, which gives the smallest estimation error covariance matrix when a linear KF is applied to a system with linear dynamic equations and with zero-mean Gaussian observation and excitation noises \cite{Simon_KF_Inequality}.
%
%
The optimization problem in \eqref{Projection_Sigma_Points} is a quadratically constrained quadratic program (QCQP), which is convex since $\W_k$ is SPD  and the constraints are convex \cite{Boyd}.
As the constraints are only on the first two elements of the state vector,
it is possible to reduce the size of the QCQP problem.
A conventional way to do so is as follows.
Suppose that $\q(1\!:\!2)$ is fixed. Then we can find the optimal $\q(3\!:\!N)$, which is a function of $\q(1\!:\!2)$.
By substituting the optimal $\q(3\!:\!N)$ into the cost function, we obtain a QCQP,
which only involves the unknown $\q(1\!:\!2)$.


However, in the above approach, we first need to find the matrix $\W_k$ through an inverse operation which is both unnecessarily
costly and numerically unstable if the covariance matrix $\boldsymbol \Sigma_{k|k}$ is ill-conditioned.
To avoid these shortcomings, we propose to use an idea from \cite{Paige} in order to reformulate and reduce the size of the convex QCQP problem in \eqref{Projection_Sigma_Points} such that it can be solved in a more numerically reliable way.
Recalling that $\boldsymbol \Sigma_{k|k} =  \U_{k|k}^{T} \U_{k|k}$, the objective function in \eqref{Projection_Sigma_Points} can be expressed as
$(\q- \s_{k|k}^{(j)})^{T} \U_{k|k}^{-1} \U_{k|k}^{-T}  (\q- \s_{k|k}^{(j)})$.
To avoid the inverse operation,
we define
\be  \label{Paige}
\u =  \U_{k|k}^{-T} ( \s_{k|k}^{(j)} - \q).
\ee
Then we have
\be \label{Paige1}
\q = \s_{k|k}^{(j)} - \U_{k|k}^T \u.
\ee
We partition the lower triangular matrix $\U_{k|k}^T$ as follows:
\be
\U_{k|k}^T =
\begin{bmatrix}
\L_{11} & \0 \\
\L_{21} & \L_{22} \\
\end{bmatrix},
\ee
where $\L_{11}\in \mathbb{R}^{2\times 2}$ and $\L_{22}\in \mathbb{R}^{(N-2) \times (N-2)}$ are lower triangular.
Then from \eqref{Paige1} we have
\be \label{eq:q12}
\q(1\!:\!2) = \s_{k|k}^{(j)}(1\!:\!2) - \L_{11} \u(1\!:\!2).
\ee
Using  \eqref{Paige} and \eqref{eq:q12}, we can reformulate the QCQP  problem   \eqref{Projection_Sigma_Points} as
\begin{align}
&  \min_{\u} \Big\lbrace \u^{T}(1\!:\!2) \u(1\!:\!2) + \u^T(3\!:\!N)\u(3\!:\!N) \Big\rbrace \label{Projection_Modified_1} .\\
& \, \mathrm{ s.t.}\   \big \|  \L_{11}\u(1\!:\!2)- (\s_{k|k}^{(j)}(1\!:\!2)  - \a^i ) \big \| \leq r_k^i+ \epsilon \sigma_n, \,  i \in {\cal N}_k. \nonumber
\end{align}
Obviously, the optimal $\u(3:N)=\0$ and the optimization problem \eqref{Projection_Modified_1} becomes
\begin{align} \label{QCQP}
&  \min_{\u} \Big\lbrace \u^{T}(1\!:\!2) \u(1\!:\!2) \Big\rbrace   . \\
& \, \mathrm{ s.t.}\  \big  \|  \L_{11}\u(1\!:\!2)- (\s_{k|k}^{(j)}(1\!:\!2)  - \a^i ) \big \| \leq r_k^i+ \epsilon \sigma_n, \,  i \in {\cal N}_k. \nonumber
\end{align}
This 2D convex QCQP problem can now be solved efficiently using iterative techniques \cite{Boyd}.

After finding the optimal $\u(1:2)$, we can compute the optimal $\q$ using \eqref{Paige1} and the fact that the optimal $\u(3\!:\!N)=\0$
as follows:
\be \label{Projected_Sigma_Points}
{\cal P}(\s_{k|k}^{(j)}) \triangleq
\q = \s_{k|k}^{(j)} -
\begin{bmatrix}
\L_{11} \\
\L_{21} \\
\end{bmatrix}
\u(1\!:\!2).
\ee
%

The above approach for reducing the size of the QCQP problem \eqref{Projection_Sigma_Points}
not only avoids a matrix inverse computation, which may cause numerical instability (see \cite{Paige}),
but it is also computationally efficient.
This approach is even more suitable when a SRUKF is employed since the Cholesky factor $\U_{k|k}$ of  $\boldsymbol \Sigma_{k|k}$ is readily provided in \eqref{Cholesky_Aposterior}.

After finding the projected sigma points through \eqref{Projected_Sigma_Points}, the mean and covariance matrix may be estimated through weighted averaging
\begin{align}\label{Mean_Sigma_Proj}
\s_{k|k}^{ {\cal P}} &= \sum_{j=0}^{2N} w^{(j)} {\cal P}(\s_{k|k}^{(j)}) , \\
\label{Cov_Sigma_Proj}
\boldsymbol \Sigma_{k|k}^{{\cal P}} &=\sum_{j=0}^{2N} w^{(j)} ({\cal P}(\s_{k|k}^{(j)}) - \s_{k|k}^{ {\cal P}}) ({\cal P}(\s_{k|k}^{(j)}) - \s_{k|k}^{{\cal P}} )^T .
\end{align}

As before, instead of \eqref{Cov_Sigma_Proj} we compute the Cholesky factor $\U_{k|k}^{{\cal P}}$ of $\boldsymbol \Sigma_{k|k}^{{\cal P}}$:
\begin{align}
& \mathbf{e}_{\cal P}^{(j)} = \sqrt{ w^{(j)} } ( {\cal P}(\s_{k|k}^{(j)}) - \s_{k|k}^{ {\cal P}}) ), \quad  j=0,\ldots, 2N , \nonumber \\
&
\U_{k|k}^{{\cal P}} = \textrm{qr} \left\lbrace  [ \mathbf{e}_{\cal P}^{(0)}, \mathbf{e}_{\cal P}^{(1)} , \ldots , \mathbf{e}_{\cal P}^{(2N)} ]^T \right\rbrace .
 \label{Mean_Cholesky_Proj_2}
\end{align}
As observed in Fig. \ref{fig:Fig1}, the projected sigma points have a different mean and covariance matrix.
The weighted average of the sigma points achieved through this technique lies inside the feasible region as the average of selected points in a convex feasible region must lie in it \cite{Linear_Nonlinear_Programming_Ye}.
Furthermore, the covariance matrix of the error is generally reduced as the sigma points have moved closer to each other.

Finally, in the next iteration of the unconstrained SRUKF, the constrained \textit{a posteriori} state estimate $\s_{k|k}^{{\cal P}}$ and the corresponding error covariance matrix $\boldsymbol \Sigma_{k|k}^{{\cal P}}$ replace $\s_{k|k}$ and $\boldsymbol \Sigma_{k|k}$, respectively as
\begin{align} 
\label{Proj_State_Update}
\s_{k|k} &= \s_{k|k}^{{\cal P}} ,\\
\boldsymbol \Sigma_{k|k} &= \boldsymbol \Sigma_{k|k}^{{\cal P}}.
\label{Proj_Covariance_Update}
\end{align}

%

\subsection{Algorithm Summary and Computational Analysis}
The proposed algorithm is summarized in Algorithm 1.
\begin{algorithm}[htbp]
\caption{CSRUKF}
\begin{algorithmic}[1]
\STATE Initialize $\s_{0|0}$ and set $\boldsymbol \Sigma_{0|0}$ to a large SPD diagonal matrix.
\STATE Set $\eta_{\alpha}$ and $\epsilon$
\FOR{$k= 1, \ldots, K$}
\STATE Prediction of $\s_{k|k-1}$ using \eqref{Prediction_Mean}, and $\U_{k|k-1}$ using \eqref{Prediction_Covariance_Cholesky}.
\IF { $|{\cal L}_k | = 0 $}
\STATE Set $\s_{k|k}=\s_{k|k-1}$  and $ \U_{k|k} =  \U_{k|k-1}$.
\ELSE
\STATE Find the predicted measurement through \eqref{Measurement_Prediction}.
\STATE Calculate the predicted mean \eqref{Measurement_Mean} and implement \textrm{qr}\{.\} in \eqref{Cholesky_Measurement_QR}.
\STATE Estimate the cross-covariance in \eqref{Cross_Covariance}.
\STATE Solve \eqref{Kalman_Gain} to find $\T_k$.
\STATE Estimate the \textit{a posteriori} mean $\s_{k|k}$ using \eqref{UKF_Aposterior} and Cholesky factor of \textit{a posteriori} covariance matrix $\U_{k|k}$ using \eqref{Cholesky_Aposterior}.
\ENDIF
\STATE Generate the sigma points using \eqref{Sigma_points_Unconstrained_2}.
\STATE For every sigma point whose first two elements fall outside ${\cal D}_k$ solve \eqref{QCQP} and find the projected point \eqref{Projected_Sigma_Points}.
\STATE Estimate $\s_{k|k}^{{\cal P}}$ using \eqref{Mean_Sigma_Proj} and $\U_{k|k}^{{\cal P}}$ using \eqref{Mean_Cholesky_Proj_2}.
\STATE Replace $\s_{k|k}^{{\cal P}}$ and $\U_{k|k}^{{\cal P}}$ as the \textit{a posteriori} estimates, i.e., \eqref{Proj_State_Update} and \eqref{Proj_Covariance_Update}.
\ENDFOR
\end{algorithmic}
\end{algorithm}

Our algorithm consists of two main parts; SRUKF and projection of sigma points, which will be discussed in more details below.

The SRUKF requires less computations compared to an ordinary UKF, thus it is more efficient.
The computational analysis of SRUKF is discussed in \cite{SRUKF}.
The main computationally demanding task in the SRUKF algorithm is finding the Cholesky factor of a matrix of size $\mathbb{R}^{N \times N}$, where $O(N^3)$ computations are needed using the QR factorization approach.

The QCQP in \eqref{QCQP} is a convex optimization problem, which is not NP hard \cite[p.153]{Boyd}, and can be solved in polynomial time using an extended optimization package in Matlab such as Sedumi \cite{SeDuMi}.
Since $\u(1\!:\!2) \in \mathbb{R}^2$, the optimization problem can be solved with low cost.
The optimization in \eqref{QCQP} has to be done for $2N+1$ sigma points at most, however, these calculations can be performed in parallel and independently of each other; hence our technique is suitable for parallel processing.
The computational cost of the algorithm depends on the number of sigma points in \eqref{Sigma_points_Unconstrained_2} that fall inside the feasible region, as the projection operation needs not to be applied on them.
By tuning the parameter $\alpha$ we achieve a trade-off between accuracy and computational cost.
If $\alpha$ is chosen to be small, then it is more likely that many sigma points will fall inside the feasible region, so the optimization problem does not need to be solved for them, resulting in a lower computational cost.
However, selecting a small $\alpha$ may degrade the localization performance as the unconstrained estimated quantities remain unchanged after applying the constraints.
On the other hand, selecting a large $\alpha$ increases the computational cost but at the same time may result in sampling many of the non-local points, and thus the linearisation of $\h(\s_k)$ might be inaccurate \cite{Julier_UT}.
In our simulations, it is observed that selecting $0.65 \leq \alpha \leq 0.85$ can offer a reasonable trade-off in terms of accuracy and computational cost.

\section{Simulation Results} \label{Sec:Simulation}
We consider a 2-D area with four fixed RNs located at known positions $\a^{1}=[0 , 0]^{T}$, $\a^{2}=[0 , 1000]^{T}$, $\a^{3}=[1000 , 1000]^{T}$, and $\a^{4}=[1000 , 0]^{T}$, where the units are in meters.
A mobile agent moves on this 2-D plane according to the motion model considered earlier in \eqref{State} with $\Q= 0.04 \I_2$ and the sample time set to $\delta t= 0.2$s for $K=1000$ time samples.

%

To model the range measurement, the true distance between each RN and MN is perturbed with a zero-mean Gaussian noise.
We consider two different measurement noise scenarios: large noise $\sigma_n=100$m and small noise $\sigma_n=10$m.
The large noise assumption can model general applications like narrowband cellular mobile positioning, while the small noise assumption is suitable for localization applications with accurate ranging, e.g., IEEE 802.15.4.a.
Note that the accuracy of UWB ranging can be improved by increasing the bandwidth of the system \cite{Gezici}.
We also perturb some of the measurements by NLOS biases which are modelled as exponential random variables with parameter $\gamma=500$m.
We consider three different scenarios in which, out of the total four measurements, the number of LOS ones $| {\cal L}_k|$ is set to be 2, 1, or 0.

For the proposed CSRUKF we consider $\epsilon=3$ (for the feasible region in \eqref{Feasible_Region_NLOS_2}) and $\alpha =70 \%$, which corresponds to $\eta_{\alpha}=4.8784$ by assuming a Gaussian posterior PDF.
Note that for CSRUKF, all the sigma points violating the constraints are projected onto the feasible region.
For solving the QCQP problem, we use the optimization toolbox \texttt{Yalmip} \cite{Yalmip} and \texttt{Sedumi} solver \cite{SeDuMi}.

In order to see if projecting all the sigma points is necessary to achieve a good result in NLOS scenarios, we first consider the common projection technique where only the  \textit{a posteriori} state estimate of a KF is projected onto the feasible region \cite{Simon_KF_Inequality}.
Therefore, $\s_{k|k}$ obtained through the SRUKF is projected onto the feasible region, thus the new \textit{a posteriori} state estimate satisfies the constraints, however, the \textit{a posteriori} estimate of the covariance matrix is not changed as compared to the unconstrained case.
This approach has in general a lower computational cost compared to the proposed CSRUKF algorithm since at most one projection operation needs to be done at each iteration.
We denote this approach by projection Kalman filter (PKF) and for solving the optimization problem we follow the similar procedure as done for CSRUKF.

For comparison purposes, we consider the conventional techniques proposed in \cite{EKF_TDOA_AOA}, \cite{EKF_NLOS_Le}, \cite{Yu_KF}, in which the range measurements are processed using a KF and then the smoothed range measurements are used in an EKF where the diagonal elements of the covariance matrix corresponding to the NLOS measurements are scaled for further mitigation of NLOS bias.
While these approaches differ slightly in terms of pre-processing and variance calculation, we consider the simple one in \cite{EKF_NLOS_Le} denoted by smooth EKF (SEKF) with scaling factor 1.5 and assume that the NLOS identification and variance calculation are done without error.

The Cramer-Rao lower bound (CRLB) analysis in NLOS shows that if no prior statistics about the distribution of the NLOS bias is available then the optimal strategy is to discard the NLOS measurements and only use LOS ones \cite{Kobayashi}.
If prior statistics are available then the NLOS measurements should also be used to achieve a lower MSE.
However, this bound can only be practical if there are enough LOS measurements for unambiguous localization, hence, for low number of LOS, i.e., less than two it can not be useful.
Even though the  posterior Cramer-Rao bound (PCRB) on positioning RMSE has been derived approximately in \cite{PCRB_NLOS_Chen, PCRB_NLOS_Fritsche}, these derivations are based on the assumption that the NLOS bias has a Gaussian distribution with known mean and variance. 
Evaluating this PCRB for other NLOS distributions such as exponential is even more challenging.
Since in this paper, there is no information about the distribution of the NLOS biases, except that they are positive, the mentioned lower bound is still loose and can not accurately show the lowest possible error in estimating the state vector.
%
%
Due to these limitations in finding a lower bound on the positioning RMSE, we consider an ideal situation where the mean and variance of the NLOS biases are known.
Then we apply a conventional EKF to the complete measurement vector $\r_k=[r_k^1, r_k^2, \ldots, r_k^M]^T$.
Therefore, the mean of the bias is subtracted from the NLOS measurements, and the covariance matrix $\R_r$ of the measurement error $\r_k$ is scaled according to the variance of the NLOS bias.
For instance if $i \in {\cal N}_k$ then $\R_r(i,i) = \sigma_n^2+ \sigma_b^2$, where $\sigma_b^2$ is the variance of the NLOS bias.
Although this approach, which is denoted by bias-aware EKF (BEKF), is not optimal when the mean and variance of the NLOS bias are known (due to the non-Gaussian errors), it can be regarded as a benchmark for comparison with our method.


To evaluate the performance of the algorithms in different scenarios, we perform $T=500$ Monte Carlo (MC) trials for each scenario and consider different trajectories at each trial.
Let $\x_k^t$ and $\x_{k|k}^{t}$ denote the true state vector and its estimate at the $k$-th time step of the trajectory over the $t$-th Monte Carlo trial, respectively.
The performance metrics are the cumulative distribution function (CDF) of positioning error $e_k$, expressed as
\be \label{CDF_Error}
\textrm{CDF}(e_k) = \mathbb{P} \Big[ (\x_{k}^t-\x_{k|k}^t)^T(\x_{k}^t-\x_{k|k}^t) \leq e_k \Big] ,
\ee
and the root mean square error (RMSE) of position estimate at time step $k$, defined as
\be \label{RMSE_Error}
\bar{\e}_k = \sqrt{\mathbb{E} \Big[ (\x_{k}^t- \x_{k|k}^{t})^T(\x_{k}^t- \x_{k|k}^{t}) \Big] },
\ee
where the quantities in \eqref{CDF_Error} and \eqref{RMSE_Error} are evaluated approximately using MC trials.

In the following, we compare the effect of measurement noise and NLOS error on the performance of different techniques in each considered scenario.
We assume that the initial estimate $\s_{0|0}$ is normally distributed with mean equal to the true state $\s_{0}$ and covariance matrix $\boldsymbol \Sigma_{0|0}= \textrm{diag}([10^4,10^4, 10^2,10^2] )$.

\begin{figure*}[htbp]
\centering
\subfloat[]{
\includegraphics[width=51mm,height=39mm]{RMSE_2LOS_Exp_Large-eps-converted-to.pdf}
}~
\subfloat[]{
\includegraphics[width=51mm,height=39mm]{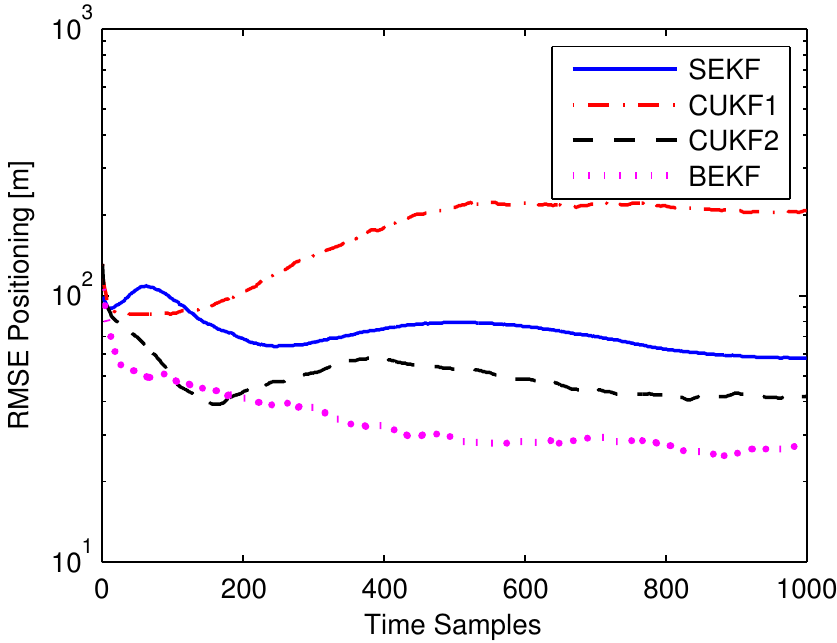}
}~
\subfloat[]{
\includegraphics[width=51mm,height=39mm]{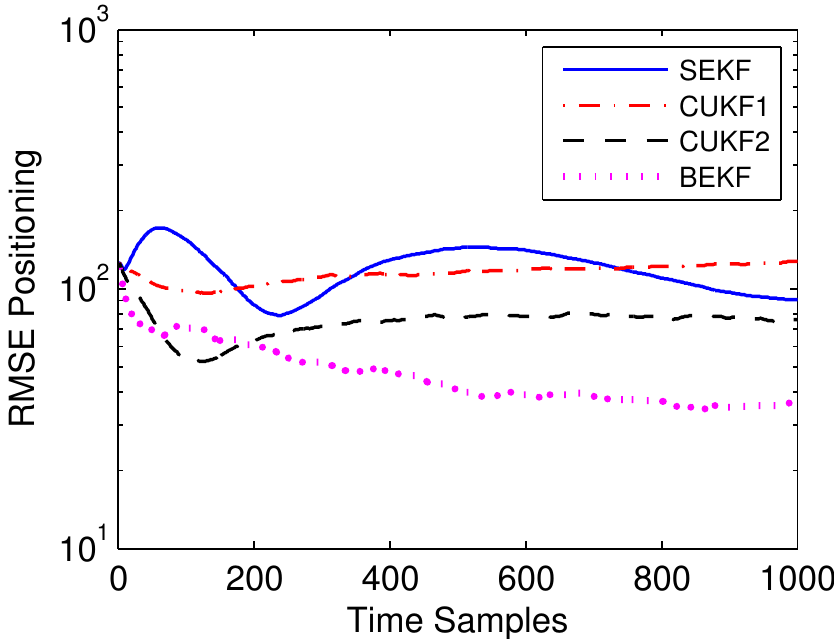}
}\\
\subfloat[]{
\includegraphics[width=51mm,height=39mm]{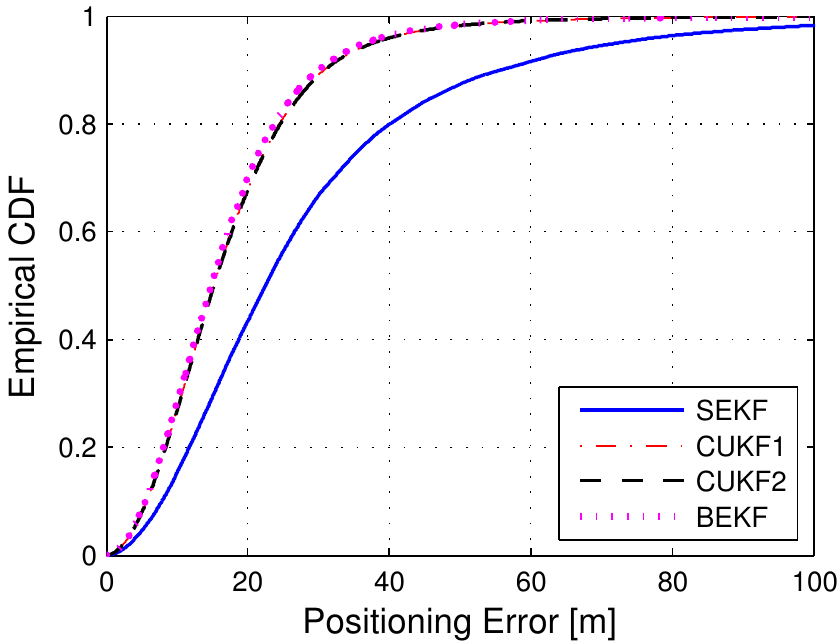}
}~
\subfloat[]{
\includegraphics[width=51mm,height=39mm]{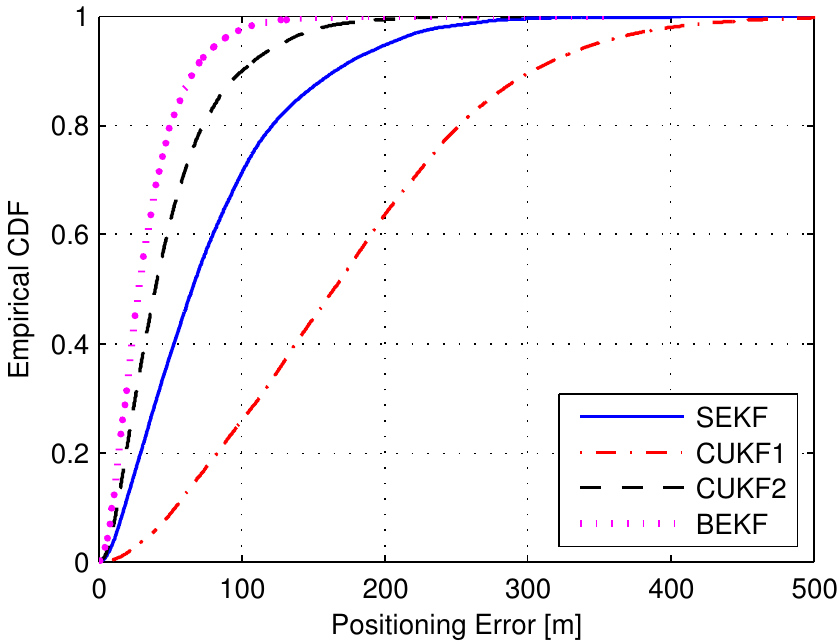}
}~
\subfloat[]{
\includegraphics[width=51mm,height=39mm]{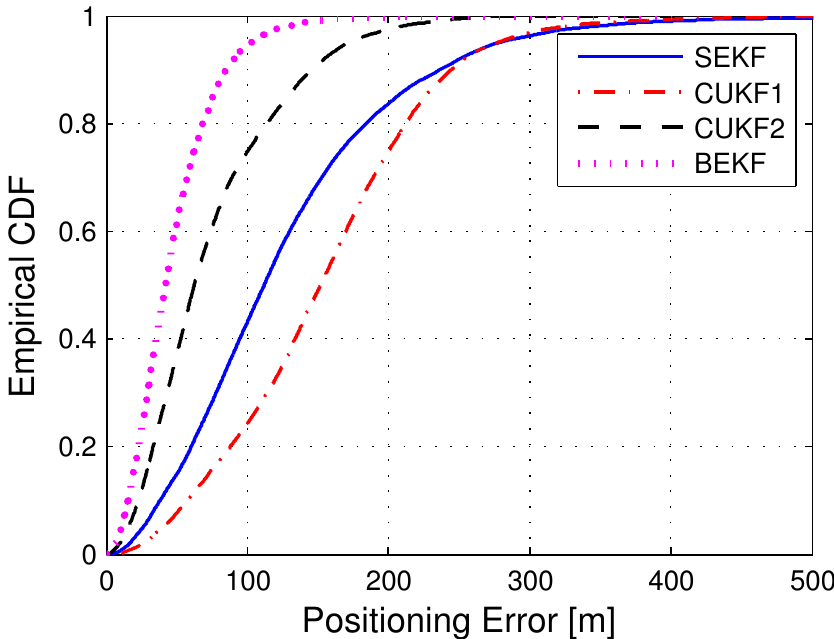}
}\\
\caption{Comparison of different techniques for large measurement noise $\sigma_n=100$m and with exponentially distributed NLOS bias with parameter $\gamma =500$m; (a) RMSE for $| {\cal L}_k|=2$, (b) RMSE for $| {\cal L}_k|=1$, (c) RMSE for $| {\cal L}_k|=0$, (d) CDF for $| {\cal L}_k|=2$, (b) CDF for $| {\cal L}_k|=1$, (c) CDF for $| {\cal L}_k|=0$. }
\label{fig:RMSE_CDF_Exp_Large}
\end{figure*}

\subsection{Large Measurement Noise}
In the first scenario, we consider the case of a narrowband ranging application where the noise variance is relatively high, i.e., $\sigma_n=100$m is considered.
The RMSE versus time step is illustrated in Fig. \ref{fig:RMSE_CDF_Exp_Large} for the scenarios $| {\cal L}_k|$ = 2,1, and 0.
The corresponding CDF of the positioning error is also plotted for each scenario.

As observed, when at least two LOS measurements are available, the RMSE of CSRUKF, PKF and that of the BEKF follow each other closely.
The RMSE of SEKF is higher than the RMSE of the other methods because the mean values of the NLOS biases are not mitigated efficiently.
The RMSE of our proposed technique is almost the same as that of an EKF with outlier rejection, however, the latter is not shown for the sake of figure visibility.
Note that for two or more LOS measurements, the EKF with outlier rejection achieves nearly optimal RMSE because the non-linearity of the range function is not very high, thus the EKF with Taylor series approximation can achieve a MSE close to the PCRB.
Therefore, our algorithm has almost the same performance as the optimal approaches in this scenario, which is a satisfactory result in terms of RMSE.
The reason is that it is quite unlikely that the sigma points in \eqref{Sigma_points_Unconstrained_2} fall outside the feasible region, thus the constraints do not change the location of the sigma points significantly.
For more number of LOS measurements, the same behaviour is observed.

If less than two LOS measurements are available, then the EKF with outlier rejection diverges due to the lack of measurements in the observation vector (see in \cite{Jourdan}).
The performance of the PKF is worse than the SEKF, which shows that by only projecting the state estimate, the performance may be poor in the absence of enough LOS measurements.
This is because the size of the feasible region is large due to the large measurement noise variance, and the first two elements of the sigma points, i.e., the ones related to the position, lie onto the boundary of the feasible region.
Note that due to the large NLOS biases, it is generally expected that the unknown position lies inside the feasible region formed by NLOS measurements rather than on its boundary.
The proposed CSRUKF, however, obtains the best performance among these approaches, and its RMSE is close to BEKF.
This shows that it is necessary to project all the sigma points in \eqref{Sigma_points_Unconstrained_2} to achieve a good performance.

\begin{figure*}[htbp]
\centering
\subfloat[]{
\includegraphics[width=51mm,height=39mm]{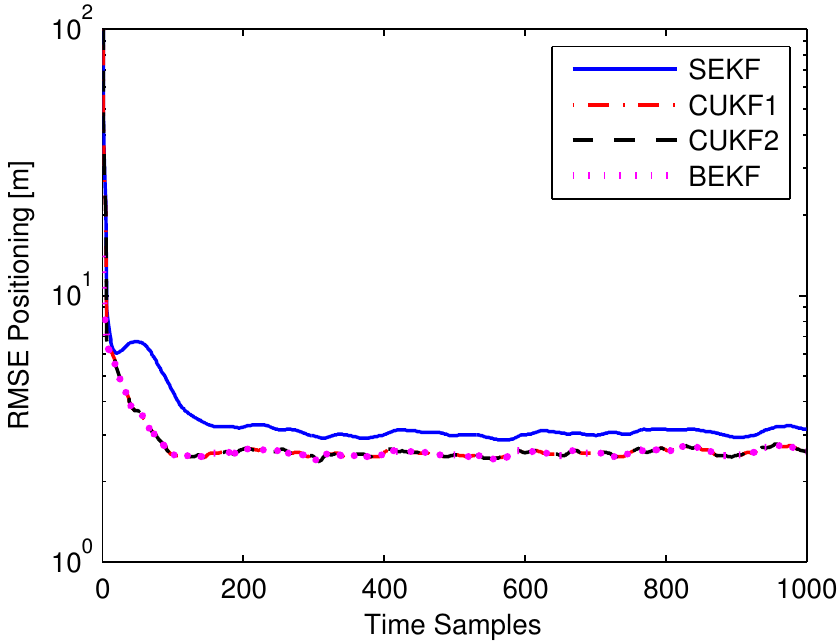}
}~
\subfloat[]{
\includegraphics[width=51mm,height=39mm]{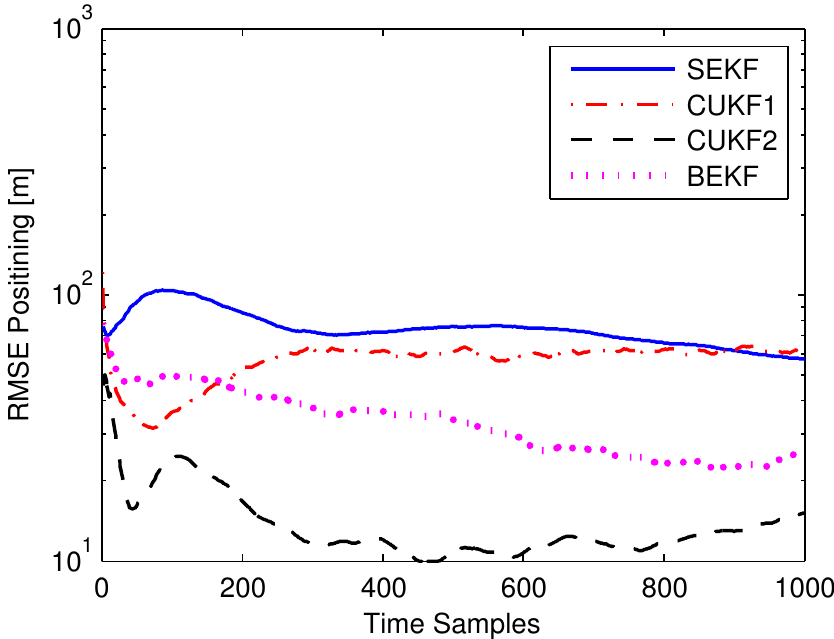}
}~
\subfloat[]{
\includegraphics[width=51mm,height=39mm]{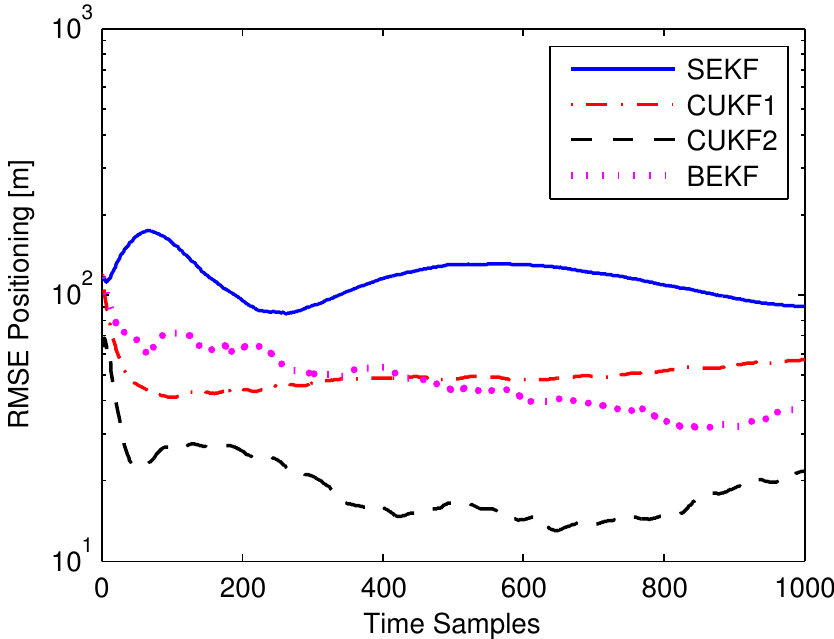}
}\\
\subfloat[]{
\includegraphics[width=51mm,height=39mm]{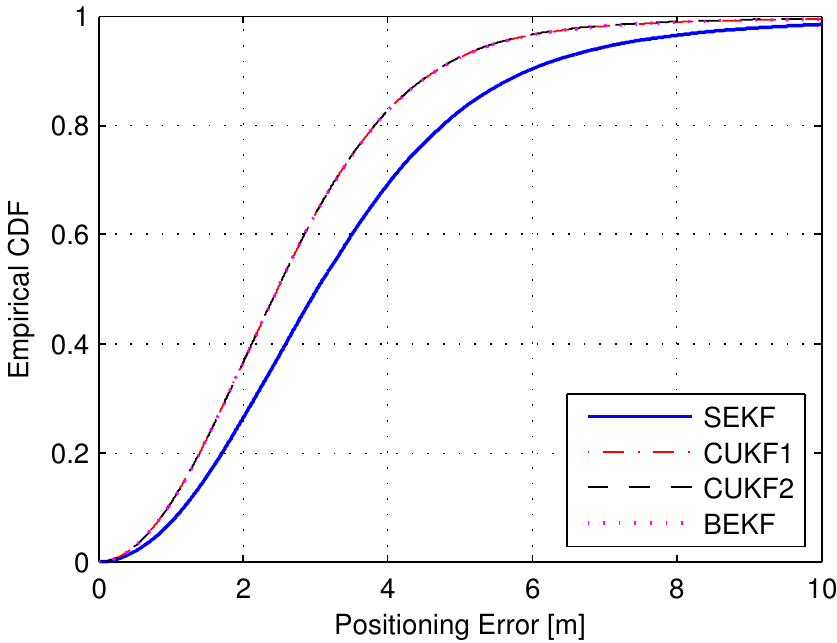}
}~
\subfloat[]{
\includegraphics[width=51mm,height=39mm]{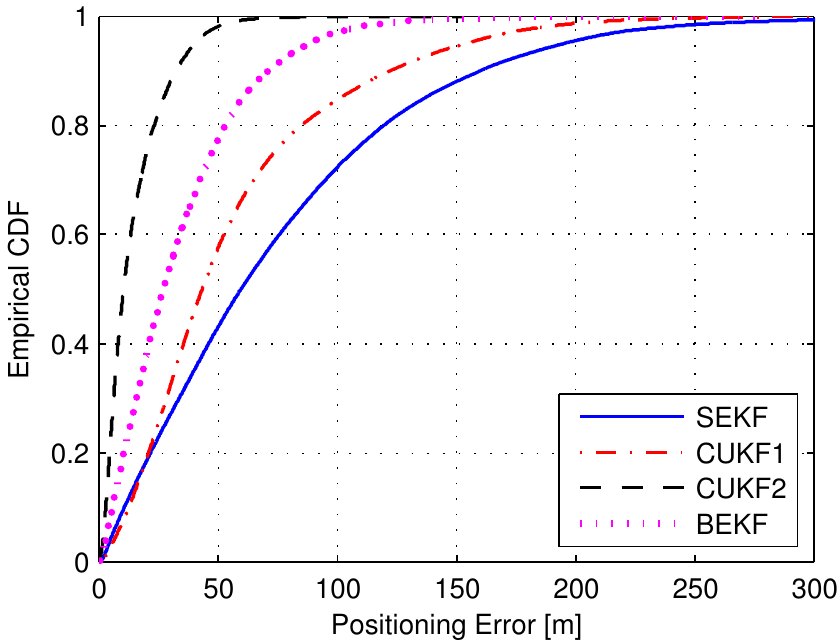}
}~
\subfloat[]{
\includegraphics[width=51mm,height=39mm]{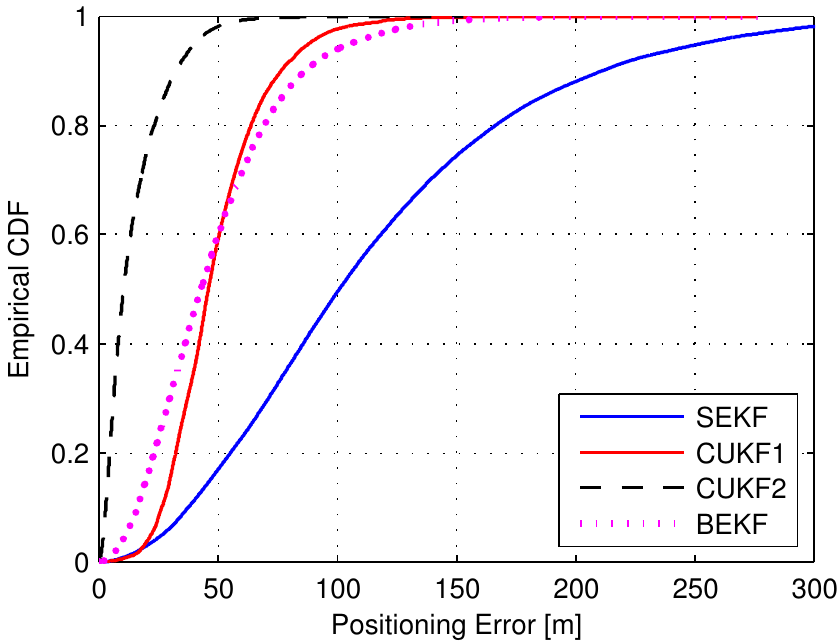}
}\\
\caption{Comparison of different techniques for small measurement noise $\sigma_n=10$m and with exponentially distributed NLOS bias with parameter $\gamma =500$m; (a) RMSE for $| {\cal L}_k|=2$, (b) RMSE for $| {\cal L}_k|=1$, (c) RMSE for $| {\cal L}_k|=0$, (d) CDF for $| {\cal L}_k|=2$, (b) CDF for $| {\cal L}_k|=1$, (c) CDF for $| {\cal L}_k|=0$. }
\label{fig:RMSE_CDF_Exp_Small}
\end{figure*}

\subsection{Small Measurement Noise}
In the second scenario, we consider a case where the noise variance is relatively small $\sigma_n=10$m, which can model the UWB signalling applications.
The RMSE is illustrated in Fig. \ref{fig:RMSE_CDF_Exp_Small} for the scenarios $| {\cal L}_k|=$ 2, 1, and 0, respectively.
The CDF is also illustrated in Fig. \ref{fig:RMSE_CDF_Exp_Small} for the considered scenarios.

In the presence of two LOS measurements, the RMSE of CSRUKF, PKF, and BEKF are close to each other as observed in Fig. \ref{fig:RMSE_CDF_Exp_Small}.
These RMSEs are close to the RMSE of an EKF with outlier rejection, although, similar to before, we have not shown the latter in this figure.
The RMSE of SEKF is slightly higher which is due to the same reason mentioned earlier.

When $|{\cal L}_k| = 1$ and  $|{\cal L}_k| = 0$, the EKF with outlier rejection diverges while the SEKF converges to a slightly lower RMSE than the initial RMSE.
The PKF has a decent performance as compared to the large noise scenario considered earlier.
This is because the size of the feasible region is smaller in this case, hence, after doing the projection of the state estimate onto the boundary, there is generally a lower uncertainty in positioning.
The performance of CSRUKF is better than PKF and SEKF with a noticeable margin.
Surprisingly, the performance of the CSRUKF is even better than BEKF, although our algorithm does not use the mean and variance of the NLOS biases, because we assume they are not available.
One of the reasons that our method performs well under small measurement noise scenario is that the size of the feasible region is small as compared to the large measurement noise, therefore, the MN position is restricted to be within a smaller area.
Hence, through processing consecutive range measurements, the sigma points can move closer and closer to each other such that their weighted average is very close to the true position.
This shows the superior performance of our algorithm in scenarios where the NLOS bias is much larger than the measurement noise, e.g., UWB ranging applications.

\subsection{Robustness to Errors in NLOS Identification}
In this part, we want to analyse the performance of our technique in the presence of NLOS identification errors, i.e., FA and missed-detection (MD), which are inevitable in some applications.

\begin{figure}[tbp]
\centering
\subfloat[]{
\includegraphics[width=68mm,height=52mm]{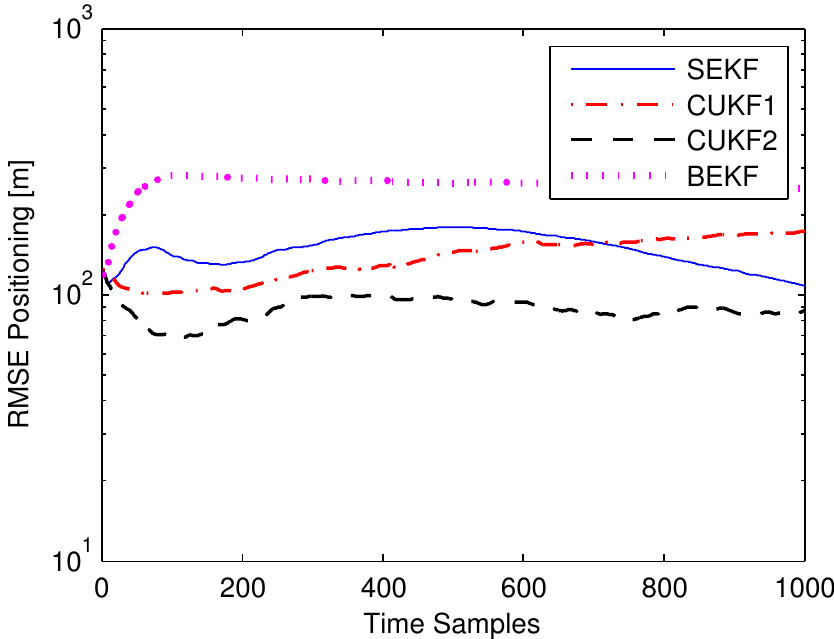}
\label{fig.:node_msd_trans_6nodes}
} \\
\subfloat[]{
\includegraphics[width=68mm,height=52mm]{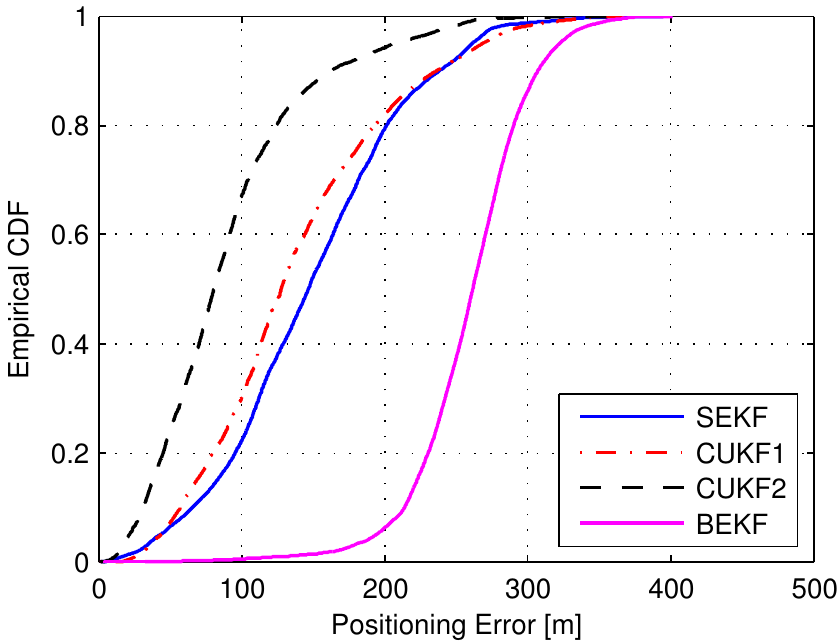}
\label{fig.:node_emse_trans_6nodes}
}
\caption{Comparison of different techniques for $\sigma_n=10$m and exponentially distributed NLOS bias with parameter $\gamma =500$m, and $| {\cal L}_k|=1$ with FA in identification; (a) RMSE, (b) CDF . }
\label{fig:FA}
\end{figure}

To see the effect of FA in NLOS identification, we assume that we have one LOS and three NLOS RNs.
However, due to the FA, the LOS link is also wrongly detected as being NLOS.
Therefore, CSRUKF, and PKF wrongly remove the LOS measurements from the measurement vector, however, they employ the wrongly detected measurement to impose a constraint on the state vector.
Since the parameter $\epsilon$ can increase the chance that a LOS measurement also satisfies the constraint in \eqref{Feasible_Region_NLOS_2}, it is expected that FA does not severely degrade the performance of our proposed technique.
The simulation results are shown in Fig. \ref{fig:FA}, where it is observed that our proposed algorithm CSRUKF is robust against FA error in NLOS identification and outperforms the SEKF.
Note that the BEKF algorithm suffers significantly from the FA NLOS identification error.

If the NLOS links are regarded as LOS ones, i.e., in the presence of NLOS MD error, all the Kalman-type filters have to use a biased measurement in their observation vector, and thus it is not surprising that their performances are degraded.
Therefore, for our algorithm to perform well in most of the times, the threshold used for NLOS identification should change such that the probability of MD becomes very small.



\section{Conclusion} \label{Sec:Conclusion}
A constrained square-root unscented Kalman filter (CSRUKF) with projection technique was considered in this paper for the aim of TOA-based localization of an MN in NLOS scenarios.
The NLOS measurements were removed from the measurement vector, instead they were employed to impose quadratic constraints onto the position coordinates of the mobile terminal.
The sigma points of the UKF which violated the constraints were projected on the feasible region by solving a convex quadratically constrained quadratic program (QCQP).
As compared to other constrained UKF techniques, we considered a square root filter and avoided computing the inverse of state covariance matrix both in the Kalman filter and in the optimization steps, thus our approach has better numerical stability and lower computational cost.
Through simulations, it was shown that, our algorithm performed better than other approaches in different NLOS scenarios.
In particular, the performance was excellent when a small measurement noise variance was considered, thus, our technique is even more suitable for high resolution TOA-based UWB localization.
Another advantage of our technique is its robustness to false alarm error in NLOS identification.
The proposed filter can be extended to the case that the information of an IMU is fused with the range measurement for more accurate mobile localization.

\ifCLASSOPTIONcaptionsoff
  \newpage
\fi



%
%
%

\bibliographystyle{IEEEtran}

\bibliography{References}

\end{document}